\documentclass[journal]{IEEEtran}
\IEEEoverridecommandlockouts
\usepackage{cite}
\usepackage{bm}
\usepackage{amsmath}
\usepackage{mathrsfs}

\usepackage{algorithmic}
\usepackage{multirow}
\usepackage{makecell}
\usepackage{array}
\ifCLASSOPTIONcompsoc
\usepackage[caption=false,font=normalsize,labelfont=sf,textfont=sf]{\part{title}subfig}
\else
\usepackage{subcaption}
\fi
\usepackage{url}
\usepackage{graphicx,amsmath,amssymb,amsfonts}
\usepackage{algorithmic,algorithm}
\usepackage{stfloats}
\usepackage{amsmath}
\allowdisplaybreaks[3]
\usepackage{xcolor}

\newtheorem{theorem}{\textbf{Theorem}}

\newtheorem{lemma}{\textbf{Lemma}}

\newtheorem{corollary}{\textbf{Corollary}}

\makeatletter

\newcommand{\Rmnum}[1]{\expandafter\@slowromancap\romannumeral #1@}

\makeatother
\begin{document}

\title{
{\fontsize{21.5 pt}{\baselineskip}\selectfont  Token Communication in the Era of Large Models: \\ An Information Bottleneck-Based Approach} 
}

\author{\IEEEauthorblockN{
		Hao~Wei, 
		Wanli~Ni,
		Wen~Wang, 
		Wenjun~Xu,~\IEEEmembership{Senior Member,~IEEE}, \\
		Dusit~Niyato,~\IEEEmembership{Fellow,~IEEE},
		and Ping~Zhang, ~\IEEEmembership{\!\!Fellow,~IEEE}
	}
	
	\thanks{H. Wei, W. Xu, and P. Zhang are with the State Key Laboratory of Networking and Switching Technology, Beijing University of Posts and Telecommunications, Beijing 100876, China. W. Xu and P. Zhang are also with the Department of Mathematics and Theories, Peng Cheng Laboratory, Shenzhen 518066, China (e-mail: hao.wei@bupt.edu.cn; wjxu@bupt.edu.cn; pzhang@bupt.edu.cn). (\emph{Corresponding author: Wenjun Xu}.)}

	\thanks{W. Ni is with the Department of Electronic Engineering, Tsinghua University, Beijing 100084, China (e-mail: niwanli@tsinghua.edu.cn).}
	
	\thanks{W. Wang is with the Pervasive Communications Center, Purple Mountain	Laboratories, Nanjing 211111, China (e-mail: wangwen@pmlabs.com.cn).}
	
	\thanks{D. Niyato is with the College of Computing and Data Science, Nanyang Technological University, Singapore 117583 (e-mail: dniyato@ntu.edu.sg).} 
}

\maketitle
\begin{abstract}
This letter proposes UniToCom, a unified token communication paradigm that treats tokens as the fundamental units for both  processing and wireless transmission.
Specifically, to enable efficient token representations,
we propose a generative information bottleneck (GenIB) principle, which facilitates the learning of tokens that preserve essential information while supporting reliable generation across multiple modalities.
By doing this, GenIB-based tokenization is conducive to improving the communication efficiency and reducing computational complexity.
Additionally, we develop $\sigma$-GenIB to address the challenges of variance
collapse in autoregressive modeling, maintaining representational diversity and stability.
Moreover, we employ a causal Transformer-based multimodal large language model (MLLM) at the receiver to unify the processing of both discrete and continuous tokens under the next-token prediction paradigm.
Simulation results validate the effectiveness and superiority of the proposed UniToCom compared to baselines under dynamic channel conditions.
By integrating token processing with MLLMs, UniToCom enables scalable and generalizable communication in favor of multimodal understanding and generation, providing a potential solution for next-generation intelligent communications.

\end{abstract}
\begin{IEEEkeywords}
Token communication, generative information bottleneck, multimodal large language model, token prediction.

\end{IEEEkeywords}

\section{Introduction}




The emergence of large foundation models such as large language models (LLMs) and multimodal large language models (MLLMs) has marked a paradigm shift in artificial intelligence, offering unprecedented capabilities in semantic understanding, reasoning, and generation\cite{MTSC}.
Beyond their success in perception and decision-making tasks, these models are beginning to reshape the landscape of communication systems by enabling more intelligent, adaptive, and content-aware interactions\cite{LLMSC}.
In this context, token communication, where information is encoded and transmitted as tokens compatible with generative models, has emerged as a new paradigm\cite{ToDMA}.
Such representations naturally align with the internal mechanisms of large models and provide a unified interface for multimodal processing and generation.
Therefore, token communication plays a vital role in bridging model-centric intelligence and enhanced transmission capability, providing a potential solution for next-generation intelligent communications in the era of large models.

Generally, transmitted information is represented as continuous high-dimensional semantic features in semantic communication (SemCom)\cite{Survey}. 
While effective in capturing high-level meanings, such representations often lack structural granularity, interpretability, and flexibility across diverse tasks and modalities.
In contrast, at the core of generative models lies the concept of the token, a meaningful processible information unit that bridges raw data and high-level semantic abstractions.
The transformation process, known as tokenization, maps segments of input signals into discrete (e.g., text and code) or continuous (e.g., image, audio, and video) latent vectors using a tokenizer. 
By decomposing complex data into structured and interpretable units, tokenization facilitates effective sequence processing and is inherently compatible with large generative models.
Particularly, the next-token prediction (NTP) paradigm
enables LLMs to capture contextual dependencies and semantic coherence and conduct sequential token generation. 
While LLMs have successfully unified understanding and generation tasks in the textual domain, recent advances show that tasks across diverse modalities can also be reformulated within NTP by transforming multimodal tokens in a unified manner\cite{NTP}. 
This highlights the potential of token-based approaches as a compact, expressive, and generalizable medium for intelligent communications across multiple modalities.

Building on this insight, token communication envisions transmitting minimal and manageable tokens over wireless channels, advancing beyond traditional bit-level transmission and high-dimensional semantic transmission.
Despite its potentials, token communication faces two core challenges.
First, it is critical to develop strategies for extracting and transmitting concise yet informative tokens under resource-constrained wireless conditions.
Another challenge lies in effectively harnessing the resulting multimodal tokens to complete both comprehension and generation tasks.

%



To overcome the above challenges, we propose a unified token communication (UniToCom) paradigm, which establishes tokens as the fundamental 
units for wireless transmission.
Drawing inspiration from the principles of compressed and informative representations\cite{IB-JSAC}, 
a generative information bottleneck (GenIB) principle is developed for tokenization.
Moreover, UniToCom integrates a causal Transformer-based MLLM \cite{Transfusion}, 
which is deployed at the base station having sufficient computing power.
The primary contributions of this letter are summarized as follows:
\begin{itemize}
	
\item We propose a novel UniToCom paradigm for multimodal transmission. 
UniToCom first transforms source data into latent tokens through tokenization. These tokens are then transmitted and interpreted by a latent MLLM for next-token prediction, whose outputs are directly recovered by de-tokenizer for task inference and content generation.

\item We propose a GenIB principle to facilitate tokenization by learning minimal sufficient latent tokens.
GenIB alleviates the issue of variance collapse in autoregressive learning and enhances the diversity and stability of token representations.
Moreover, we introduce an MLLM to perceive and generate multimodal tokens in a unified manner.

\item Simulation results demonstrate the superiority of UniToCom under dynamic channel conditions. The proposed GenIB-based tokenization and the integration of MLLM boost both multimodal understanding and generation performance.
Besides, moderate token compression improves both communication and computational efficiency.

\end{itemize}

\section{Prpoposed UniToCom Framework}
\begin{figure*}[t!]
	\centering
	\includegraphics[width=0.95\textwidth]{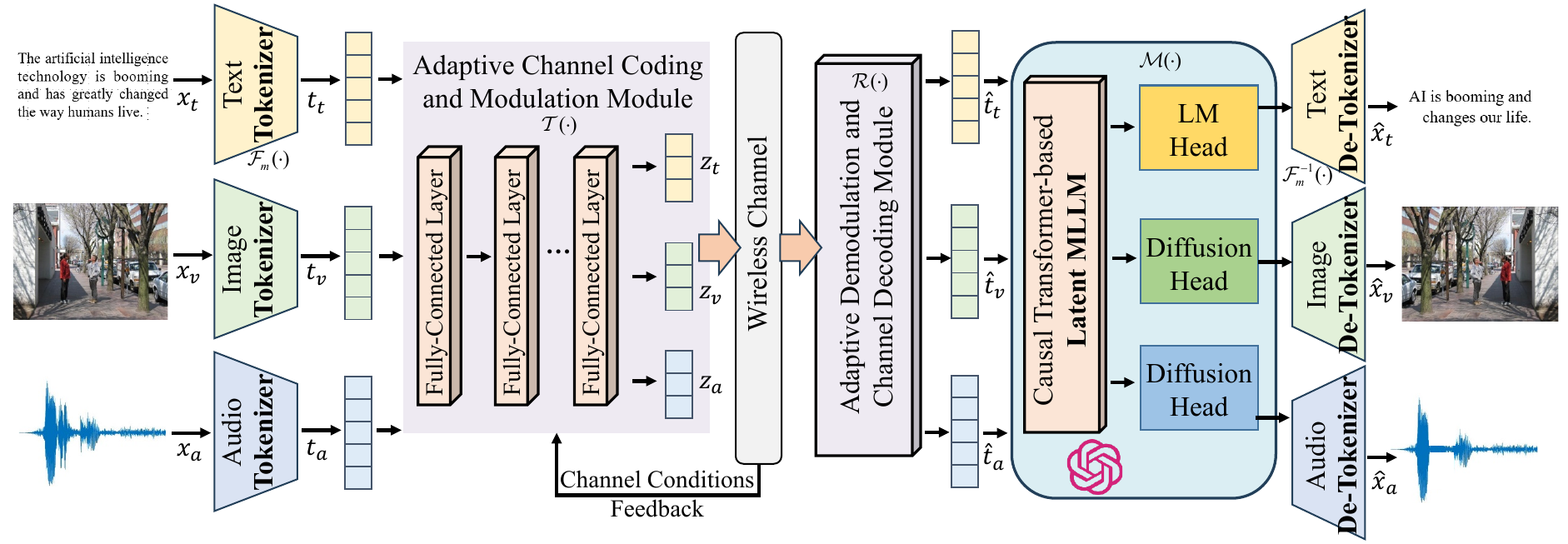}
	\caption{The proposed UniToCom paradigm: Processing and transmitting tokens for multimodal understanding and generation.}
	\label{model}
\end{figure*}

As illustrated in Fig. \ref{model}, we consider a wireless communication network, where three data modalities are considered, i.e., text ($t$), image ($v$), and audio ($a$).
The transceiver design of the proposed UniToCom is provided as follows.

\subsection{UniToCom Transmitter}
At the UniToCom transmitter, each user comprises a tokenizer and an adaptive channel coding-modulation module.
Specifically, various input modalities are first fed into the corresponding tokenizer to compress the source data $\boldsymbol x_{m}$ into latent tokens $\boldsymbol t_{m}$.
Afterwards, the extracted tokens undergo joint channel coding and modulation by the adaptive coding-modulation module\cite{SWinJSCC} to obtain the transmitted signals $\boldsymbol z_{m}$.
The encoding process is formulated as
\begin{eqnarray} 
	\boldsymbol z_{m}={\mathcal {T}(\mathcal F_{m}(\boldsymbol x_{m}))}, ~~ m\in\{t,v,a\},
\end{eqnarray}
where ${\mathcal {T}(\cdot)}$ is the adaptive channel coding-modulation module for processing modality $m$, $\mathcal F_{m}(\cdot)$ denotes the tokenizer,
$\boldsymbol x_{m}\in\mathbb{R}^{{L}_m}$ is the input modality $m$ with length ${L}_m$, $\boldsymbol t_{m}=\mathcal F_{m}(\boldsymbol x_{m})\in\mathbb{R}^{{S}_m}$ is the token with length ${S}_m$, and $\boldsymbol z_{m}\in\mathbb{C}^{{N}_m}$ is the encoded signal with dimension ${N}_m$.
Before transmitting $\boldsymbol z_{m}$ into the wireless channel, a power constraint $P$ is carried out on $\boldsymbol z_{m}$, i.e., $\frac{1}{N_m} \mathbb{E}\left\|\boldsymbol z_m\right\|^2 \leq P$.

\subsection{UniToCom Receiver}
Then, $\boldsymbol z_{m}$ is sent through a wireless channel, given by
\begin{eqnarray} 
	\boldsymbol {\hat z}_{m}=h_m\boldsymbol z_m+ \boldsymbol n_m,  ~~m\in\{t,v,a\},
\end{eqnarray}
where $h_m\in\mathbb{C}$ is the channel gain coefficient and $\boldsymbol n_{m}\sim\mathcal{CN}(\mathbf{0},\;\sigma_n^2\mathbf{I})$ denotes the additive white Gaussian noise.

At the UniToCom receiver, the received signals $\boldsymbol {\hat z}_{m}$ are first input to the adaptive demodulation and channel decoding module\cite{SWinJSCC} to reconstruct the received tokens $\boldsymbol {\hat t}_{m}$, which are then fed into the latent MLLM to perceive and generate multimodal tokens in a unified way. Finally, the de-tokenizer produces the final outputs $\boldsymbol {\hat x}_m$ by leveraging the multimodal token sequences.
The decoding process is given by
\begin{eqnarray} 
	\boldsymbol {\hat x}_m=\mathcal F_{m}^{-1}(\mathcal{M}_m({\mathcal {R}(\boldsymbol {\hat z}_{t})},{\mathcal {R}(\boldsymbol {\hat z}_{v})},{\mathcal {R}(\boldsymbol {\hat z}_{a})})),
\end{eqnarray}
where $\mathcal {\mathcal {R}(\cdot)}$ is the adaptive demodulation and channel decoding module, $\mathcal{M}_m(\cdot)$ denotes the MLLM model to generate tokens for modality $m$, and $\mathcal F_{m}^{-1}(\cdot)$ is the de-tokenizer. 

\section{GenIB-Guided Tokenization}


\subsection{GenIB Problem}

For notation simplicity, $\mathcal F_{{\boldsymbol{\alpha}}}(\cdot)$ and $\mathcal F_{{\boldsymbol{\beta}}}(\cdot)^{-1}$ are denoted as the tokenizer and de-tokenizer parameterized by $\boldsymbol{\alpha}$ and $\boldsymbol{\beta}$, respectively.
Given the defined model, 
the variables are characterized by the following probabilistic chain:
$\small{X \xrightarrow {\mathcal F_{{\boldsymbol{\alpha}}}(\cdot)} T \rightarrow {\hat T} \xrightarrow {\mathcal F_{{\boldsymbol{\beta}}}(\cdot)^{-1}} \hat{X}}$.
Notably, traditional IB approaches are primarily designed for supervised learning scenarios in a task-oriented manner, relying on labeled data to extract task-relevant features, and are thus unsuitable for source restoration and generation that do not require label transmission\cite{CL-Semcom}. 
In this work, we propose the GenIB principle, which aims to 
extract concise latent token representation $T$, as well as preserving sufficient informative information to generate $X$ as much as possible.
Such a principle can be deemed as a \emph{rate-distortion} tradeoff between the rate (i.e., compressed latent token size) and the distortion of the reconstruction/generation goal,
formulated by
\begin{eqnarray}
	\label{P0_IB}
	\underset{p(\boldsymbol {t}|\boldsymbol x)\in\mathbb{P}}{\min}&\!\!\!\!{I}(X; T)\ \ 
	\operatorname{s.t.}&\!\!\!\!{I}(\hat{T}; X)\geqslant \chi,
\end{eqnarray}
where $\mathbb{P}$ is the set of all possible probabilistic mappings, and $\chi$ represents the minimum generative informativeness level. 
Typically, the GenIB problem can be addressed using the Lagrange multiplier method, which is employed to find the minimum of a multivariable function, given by
\begin{eqnarray}
	\label{P1_IB}
	\mathcal{L}_{\text {GenIB}}\!\!\!\!\!\!\!\!&{}&={\xi} \underbrace{I(X; T)}_{\text {Rate}}\underbrace{-I({\hat T};X)}_{\text{Distortion}},
\end{eqnarray}
where ${\xi}>0$ controls the tradeoff between compressing and generation.
A smaller value of ${\xi}$ indicates a greater emphasis on informativeness with a better reconstruction/generation quality, at the expense of reduced compression efficiency.

\subsection{Variational GenIB}

For the rate term in (\ref{P1_IB}), the mutual information between the input $X$ and the latent token representation $T$ is given by
\begin{subequations}
	\begin{eqnarray}
		\label{rate0}
		I(T, X)\!\!\!\!\!&\!\!\!\!\!\!\!\!\!\!\!\!\!\!\!\!\!\!\!\!\!\!\!\!\!\!\!\!\!\!\!\!\!\!\!= \int p(\boldsymbol x, \boldsymbol {t}) \log \frac{p_{\boldsymbol{\alpha}}(\boldsymbol {t} | \boldsymbol x)}{p(\boldsymbol {t})} d \boldsymbol {t} d \boldsymbol x \\ 
		&\!\!\!\!\!\!\!\!\!\!\!\!\!\!=\int p(\boldsymbol x, \boldsymbol {t}) \log p_{\boldsymbol{\alpha}}(\boldsymbol {t} | \boldsymbol x) d \boldsymbol {t} d \boldsymbol x \! -\! \int p(\boldsymbol {t})\log p(\boldsymbol {t}) d \boldsymbol {t}. 
	\end{eqnarray}
\end{subequations}
However, the marginal distribution, defined as $p(\boldsymbol{t})=\int {p_{\boldsymbol{\alpha}}(\boldsymbol{t}|\boldsymbol x)p(\boldsymbol{x})} d\boldsymbol x$ is often intractable.
To overcome this, we apply a variational prior distribution $q(\boldsymbol{t})$ to approximate $p(\boldsymbol{t})$\cite{VIB}.
Due to the non-negative property of Kullback-Leibler (KL) divergence, it can be obtained that $D_{KL}{(p(\boldsymbol { t})||q({\boldsymbol {t}}))}\geqslant 0$,
and thus we have
$\int p(\boldsymbol { t})\log p(\boldsymbol {t}) d \boldsymbol { t}\geqslant \int  p(\boldsymbol { t})\log q(\boldsymbol { t}) d \boldsymbol {t}$.
Then, the variational upper bound is represented as
\begin{subequations}
	\begin{eqnarray}
		\label{rate1}
		I(T, X) \leqslant&\!\!\! \int p(\boldsymbol x) p_{\boldsymbol{\alpha}}(\boldsymbol {t} | \boldsymbol x) \log \frac{p_{\boldsymbol{\alpha}}(\boldsymbol {t} | \boldsymbol x)}{q(\boldsymbol {t})}d \boldsymbol x d \boldsymbol {t} \\
		=&\!\!\!\!\!\!\!\!\!\!\!\!\!\!\!\!\!\!\!\!\!\!\!\!\!\!\! D_{KL}{(p_{\boldsymbol{\alpha}}(\boldsymbol {t} |\boldsymbol x)||q({\boldsymbol {t}}))}.
	\end{eqnarray}
\end{subequations}
Generally, the prior distribution is set as $q({\boldsymbol {t}})=\mathcal{N}(\boldsymbol {t}|0,\mathbf{I})$ to induce a certain degree of sparsity, which helps to reduce the dimensions of latent vectors and mitigate the computational complexity.
Thus the variational bound is recast as
\begin{eqnarray}
	\label{rate2}
	I(T, X) \leqslant D_{KL}{\left(p_{\boldsymbol{\alpha}}(\boldsymbol {t} |\boldsymbol x)||\mathcal{N}(\boldsymbol {t}|0,\mathbf{I})\right)}.
\end{eqnarray}

Next, the distortion term $I(\hat T, X)$ is considered as
\begin{subequations}
\begin{eqnarray}
	I(\hat T, X)=&\!\!\!\!\!\!\!\!\!\!\!\!\!\!\!\!\!\!\!\!\!\!\!\!\!\!\!\!\!\!\!\!\!\!\!\!\!\!\!\! -H(X|\hat T) +H(X) 
	\\	\label{distortion0}
	\geqslant&\!\!\!\!  \int p(\boldsymbol { t}, \boldsymbol { x}) \log {p(\boldsymbol {x} | \boldsymbol { t})} d \boldsymbol { t} d \boldsymbol x +H(X) ,
\end{eqnarray}
\end{subequations}
where (\ref{distortion0}) holds due to the data processing inequality.
However, it is usually spiny to calculate the posterior distribution $p(\boldsymbol x|\boldsymbol{t})$, which is formulated as $p(\boldsymbol x|\boldsymbol{t})
	={p(\boldsymbol{ {t}}|\boldsymbol x)}{p({\boldsymbol{x}})}/p(\boldsymbol{{ t}})$.
To tackle this problem, we leverage the variational distribution $q_{\boldsymbol{\beta}}(\boldsymbol x|\boldsymbol{{ t}})$ to approximate the true posterior $p(\boldsymbol x|\boldsymbol{{t}})$.
Since $D_{KL}{(p(\boldsymbol x|\boldsymbol{{ t}})||q_{\boldsymbol{\beta}}(\boldsymbol x|\boldsymbol{{t}}))}\geqslant 0$, it can be obtained that 
\begin{eqnarray}
	&\int  p(\boldsymbol x|\boldsymbol{{ t}}) \log p(\boldsymbol x|\boldsymbol{{ t}}) d \boldsymbol x\geq \int  p(\boldsymbol x|\boldsymbol{{ t}}) \log q_{\boldsymbol{\beta}}(\boldsymbol x|\boldsymbol{{ t}}) d \boldsymbol x,
\end{eqnarray}
and thus the variational lower bound is denoted as 
\begin{subequations}
	\begin{eqnarray}
		I(\hat T,X)  \geqslant&\!\!\!\! \int p(\boldsymbol { t}, \boldsymbol x) \log {q_{\boldsymbol{\beta}}(\boldsymbol x|\boldsymbol{{t}})} d \boldsymbol {t} d \boldsymbol x +H(X) \\ 	\label{distortion1}
		\equiv &\!\!\!\!\!\!\!\!\!\!\!\!\!\!\!\!\!\!\!\!\!\!\!\!\int p(\boldsymbol { t}, \boldsymbol x) \log {q_{\boldsymbol{\beta}}(\boldsymbol x|\boldsymbol{{ t}})} d \boldsymbol { t} d \boldsymbol x,
	\end{eqnarray}
\end{subequations}
where the entropy $H(X)$ is a constant that can be neglected. 
By integrating (\ref{rate2}) and (\ref{distortion1}) into (\ref{P1_IB}), the objective function of tokenization is formalized as
\begin{eqnarray}
	\nonumber
	\mathcal{L}_{\text{GenIB}}({\boldsymbol{\alpha},\boldsymbol{\beta}})=&\!\!\!\!\!\!\!\!\!\!\!\xi D_{KL}{(p_{\boldsymbol{\alpha}}(\boldsymbol {t} |\boldsymbol x)||\mathcal{N}(\boldsymbol {t}|0,\mathbf{I}))}\\ 	\label{GIB}
	&-\int p(\boldsymbol { t}, \boldsymbol x) \log {q_{\boldsymbol{\beta}}(\boldsymbol x|\boldsymbol{{ t}})} d \boldsymbol { t} d \boldsymbol x.
\end{eqnarray}
Note that the conditional entropy in (\ref{GIB}) is commonly solved by the 
mean-square-error and cross-entropy loss.

\begin{algorithm}[t]
	\caption{$\sigma$-GenIB-Based Model Training}
	\label{algorithm}
	\textbf{Input:} Training data, epoch $C$, sample times $K$, learning rate $\eta$, modality $m$, channel noise $\boldsymbol n$.
	\begin{algorithmic}[1]
		\STATE Initialize parameters $\boldsymbol {\alpha}_m^0$ and $\boldsymbol {\beta}_m^0$ randomly.
		\STATE Encapsulate $\mathbf{w}_m^0=(\boldsymbol {\alpha}_m^0,\boldsymbol {\beta}_m^0)$.
		\WHILE{training epoch $j=1$ to $C$}
		\STATE $\boldsymbol t_{m} \longleftarrow{\mathcal F_{{\boldsymbol{\alpha}}}(\boldsymbol x_m)}$;
		\STATE Calculate the estimated KL divergence based on (\ref{rate2});
		\STATE Sample the stochastic features $\left\{\boldsymbol{t}_m\right\}_{k=1}^K \sim p(\sigma^2)$;
		\STATE Perform adaptive coding-modulation to obtain ${\boldsymbol{z}_m}$;
		\STATE Transmit ${\boldsymbol{z}_m}$ over the wireless channel, and perform demodulation and decoding to obtain ${\boldsymbol {\hat t}_m}$;
		\STATE $\boldsymbol {\hat x}_m \longleftarrow{\mathcal F_{{\boldsymbol{\beta}}}^{-1}(\boldsymbol {\hat t}_m)}$;
		\STATE Calculate the distortion term based on (\ref{Monte}); 
		\STATE Calculate the total loss based on (\ref{GIB1});
		\STATE Update the tokenizer and de-tokenizer via backward propagation with stochastic gradient descent:
		$\mathbf{w}_m^{(j)}=\mathbf{w}_m^{(j-1)}-\eta \nabla_{\mathbf{w}_m^{(j-1)}} \mathcal{L}_{\text {tok}}\left(\mathbf{w}_m^{(j-1)}\right) $;
		\STATE $j\longleftarrow j+1$;
		\ENDWHILE
	\end{algorithmic}
	\textbf{Output:} The converged parameters $\mathbf{w}_m^{(C)}=(\boldsymbol \alpha_m^{(C)},\boldsymbol \beta_m^{(C)})$.
\end{algorithm}

Particularly, autoregressive generation inherently brings in sampling uncertainty, making the variance of latent representations a critical factor.
However, in vanilla IB and variational auto-encoders (VAEs), certain latent dimensions often suffer from variance collapse, which degrades the effectiveness of autoregressive modeling.
Besides, the existing generative models tend to rely solely on stochastic representations, leading to the unstable performance.
Therefore, we propose $\sigma$-GenIB to mitigate variance collapse and improve robustness by imposing a fixed variance constraint and integrating deterministic representation into the stochastic latent space.
Specifically, the multivariate Gaussian distribution is employed to characterize 
$p_{\boldsymbol{\alpha}}(\boldsymbol {t} |\boldsymbol x)$, i.e., $p_{\boldsymbol{\alpha}}(\boldsymbol {t} |\boldsymbol x)=\mathcal{N}(\boldsymbol{ t}|\boldsymbol{\mu}, \boldsymbol{\sigma})$,
where $\boldsymbol{\mu}=\mathcal F_{{\boldsymbol{\alpha}}}(\boldsymbol{x})$  is the learned mean vector, and $\boldsymbol{\sigma}$ is the fixed variance in the latent space.
The learned mean vector $\boldsymbol{\mu}$ can be deemed as the deterministic representation, while $\boldsymbol{t}$ is regarded as the latent token representation with uncertainty, denoted as $
\boldsymbol{t}=\boldsymbol{\mu}+\boldsymbol{\sigma}\odot\boldsymbol\epsilon,  \boldsymbol\epsilon\sim\mathcal{N}(\mathbf{0},\mathbf{I}), \boldsymbol\sigma\sim\mathcal{N}(\mathbf{0},\boldsymbol{C}_\sigma)$,
where $\boldsymbol{C}_\sigma$ is a hyperparameter.
By considering both deterministic and stochastic representations, the loss function is expressed as  
\begin{eqnarray}
\nonumber
	\mathcal{L}_{{\sigma\text{-GenIB}}}({\boldsymbol{\alpha},\boldsymbol{\beta}})\!\!\!\!\!\!\!\!\!\!\!\!\!\!\!\!\!\!\!\!\!\!&=\xi D_{KL}{(\mathcal{N}(\boldsymbol{ t}|\boldsymbol{\mu}, \boldsymbol{\sigma})||\mathcal{N}(\boldsymbol {t}|0,\mathbf{I}))} \\ 	 \label{GIB1}
	&\!\!\!\!\!\!\!\!\!\!\!\!\!\!\!\!+ \lambda \mathbb{E}_{p_{\boldsymbol{\alpha}}(\boldsymbol {t} |\boldsymbol x)}\mathrm{CE}\left( \mathcal F_{{\boldsymbol{\beta}}}(\boldsymbol t), \boldsymbol{x}\right)+(1-\lambda)\mathrm{CE}\left( \mathcal F_{{\boldsymbol{\beta}}}(\boldsymbol \mu), \boldsymbol{x}\right),
\end{eqnarray}
where the weight $\lambda$ governs the tradeoff between deterministic and stochastic representations.
To enable gradient-based optimization, the reparameterization technique with Monte Carlo sampling is employed to facilitate back-propagation,
and hence the expectation term in (\ref{GIB1}) is calculated by
\begin{eqnarray}
	\label{Monte}
\mathbb{E}_{p_{\boldsymbol{\alpha}}(\boldsymbol {t} |\boldsymbol x)}\mathrm{CE}\left(\mathcal F_{{\boldsymbol{\beta}}}( \boldsymbol{{ t}}), \boldsymbol{x}\right) \approx \frac{1}{K}\sum\nolimits_{k=1}^{K}{\mathrm{CE}\left( \mathcal F_{{\boldsymbol{\beta}}}( \boldsymbol {t}_k), \boldsymbol{x}\right)},
\end{eqnarray}
where $K$ is the sample times. 
%
The detailed training process of the GenIB-based tokenization is summarized in Algorithm \ref{algorithm}.

\section{MLLM-based Next Token Prediction}
At the receiver, we leverage a causal Transformer-based latent MLLM to perceive and generate multimodal tokens 
in a unified way. 
Discrete tokens, such as text and code, are generated through next-token prediction using language modeling. Continuous tokens (e.g., image, audio, and video) are generated via next-token diffusion.

\subsection{Language Modeling}
Let $\boldsymbol{\hat t^d}=[\boldsymbol d_1,\ldots,\boldsymbol d_N]$ and $\boldsymbol{\hat t^c}=[\boldsymbol c_1,\ldots,\boldsymbol c_M]$ denote the input sequences of $N$ discrete tokens and $M$ continuous tokens recovered by the adaptive demodulation-decoding module, respectively.
These tokens are packed into $\boldsymbol{T}^0=[\boldsymbol{\hat t^d},\boldsymbol{\hat t^c}] \in\mathbb{R}^{(N+M)\times D}$ where $D$ is the hidden dimension size.
Then, $\boldsymbol{T}^0$ is fed into the causal Transformer-based latent MLLM.
Specifically, the latent MLLM consists of $L$ stacked Transformer layers, utilizing causal masking to facilitate autogressive generation.
The initial input $\boldsymbol{T}^0$ is progressively contextualized to produce the output $\boldsymbol{T}^L$, where $\boldsymbol{T}^L=\text{Transformer}(\boldsymbol{T}^{L-1})$ for $l\in[1,L]$.
The final outputs $[\boldsymbol h,\ldots,\boldsymbol h_{N+M}]=\text{Norm}(\boldsymbol T^L)$ are subsequently used to decode the predicted tokens.
For discrete tokens, the language modeling predicts the probability of the token sequence $P_d(\boldsymbol{\hat t^d})$.
The prediction process of the language modeling head is expressed as
\begin{subequations}
\begin{eqnarray}
\text{Transformer}(\boldsymbol d_i|\boldsymbol d_{<i})=&\!\!\!\!\!\!\!\!\!\!\!\!\! \!\!\!\! \!\text{Sample}(P_d(\boldsymbol d_i|\boldsymbol d_{<i})) \\
=&\!\!\!\!\text{Sample}(\text{softmax}(\boldsymbol h_i \boldsymbol W_v)),
\end{eqnarray}
\end{subequations}
where $\boldsymbol W_v\in\mathbb{R}^{D\times V}$ is the weight of the softmax classifier with $V$ being the vocabulary size, and $\text{Sample}(\cdot)$ is a sampling operation such as greedy decoding and top-$p$ sampling. This defines an autogressive classification task, in which the probability distribution of each discrete token $\boldsymbol d_i$ is predicted conditionally based on the preceding sequence $\boldsymbol d_{<i}$ using $P_d$.
The objective of the language modeling is expressed as
\begin{eqnarray}
\mathcal{L}_{\text{LM}}=-\sum\nolimits_{\boldsymbol{d}_i}\log P_d(\boldsymbol d_i|\boldsymbol d_{<i}).
\end{eqnarray}

\subsection{Next-Token Diffusion}

For continuous tokens, denoising diffusion probabilistic models (DDPM) is used in the latent MLLM. The diffusion head iteratively refines and generates each latent token $\boldsymbol c_i$, conditioned on the hidden state $\boldsymbol{h}_i$. The predicted $\boldsymbol c_i$ is then fed into the subsequent Transformer layer. The process is 
\begin{eqnarray}
\text{Transformer}(\boldsymbol c_i|\boldsymbol c_{<i})=\text{Diffusion}(\boldsymbol{h}_i).
\end{eqnarray}
Diffusion is formulated as a two-stage process: the forward process progressively adds noise to the input data, while the reverse process is learned to iteratively denoise the corrupted inputs, thereby reconstructing the original signal.
\subsubsection{Forward Process}
Given a data point $\boldsymbol c_i^0=\boldsymbol c_i$, noise is successively added over $R$ steps, with the noisy representation at step $r$ indicated by $\boldsymbol c_i^r$, where $r\in\{1,\ldots,R\}$. The noise injection follows a Markov process, defined as $q\left(\boldsymbol c_i^r | \boldsymbol c_i^{r-1}\right)=\mathcal{N}\left(\boldsymbol c_i^r | \sqrt{1-\phi_r} \boldsymbol c_i^{r-1}, \phi_r \mathbf{I}\right)$, where $\phi_r$ is a predefined noise schedule.
This forward diffusion process is reparameterized to allow direct sampling of $\boldsymbol c_i^r$ from $\boldsymbol c_i^0$, given by
\begin{eqnarray}
\boldsymbol c_i^r= \sqrt{\omega_r} \boldsymbol c_i^0+\sqrt{1-\omega_r}\boldsymbol\epsilon ,
\end{eqnarray}
where $\omega_r=\prod_{i=1}^r\left(1-\phi_i\right)$.

\subsubsection{Reverse Process}
The diffusion model is trained to approximate the reverse process $p\left(\boldsymbol c_i^{r-1} | \boldsymbol c_i^{r}, \boldsymbol h_i\right)$, which iteratively denoises the input data.
Specifically, DDPM learns a model $\boldsymbol\epsilon_\psi(\boldsymbol c_i^{r},r,\boldsymbol h_i)$ to estimate the noise $\boldsymbol\epsilon$ presented in $\boldsymbol c_i^{r}$ at diffusion step $r$.
The model parameters are optimized by minimizing the following loss:
\begin{eqnarray}
\mathcal{L}_{\text {Diff }}=\mathbb{E}_{\boldsymbol c_i, r, \boldsymbol\epsilon}\left\|\boldsymbol\epsilon-\boldsymbol\epsilon_\psi\left(\boldsymbol c_i^{r},r, \boldsymbol h_i\right)\right\|^2.
\end{eqnarray}


\section{Simulation Results}

To evaluate the performance of UniToCom, we conduct experiments on three widely-used understanding and generation tasks.
Specifically, the CLEVR dataset is employed for the visual question answering (VQA) task, the MS-COCO dataset is used for text-to-image generation, 
and the LibriSpeech dataset is adopted for automatic speech recognition(ASR).
The evaluation metrics are selected: answer accuracy for VQA, Frechet Inception distance (FID) score for text-to-image generation, and word error rate (WER) for ASR.
The proposed $\sigma$-GenIB-based tokenizer and de-tokenizer are trained with 200 epochs. Both of them have 6 Transformer layers. The batch size is 256 and the optimizer is AdamW with a learning rate of $2\times10^{-4}$. The weight decay is set to 0.01. Input image size is $256\times256$ and audio data is resampled to 24 kHz. 
The pretrained adaptive channel coding and modulation module follows the same configuration as that in \cite{SWinJSCC}. Additionally, a pretrained Transformer-based MLLM is used as the backbone with a model size of 1.4B, a hidden size of 2048, and 24 layers\cite{Transfusion}.
For performance comparison, the following baselines are considered:
1) UniToCom+VAE: This baseline adopts a VAE-based tokenizer and de-tokenizer, while the remaining modules mirror the architecture of UniToCom;
2) Vanilla SemCom: This setup utilizes a basic SemCom framework to explore and transmit semantic features, i.e., MU-DeepSC\cite{MU-DeepSC}, GenSemCom\cite{GenSemCom}, and DeepSC-ST\cite{DeepSC-ST}; 
3) Traditional Scheme: 
UTF-8, JPEG, and AMR-WB are used for text, image, and audio source coding, respectively. For channel coding, LDPC is employed for image and audio, and Turbo codes for text. The modulation scheme is set to 16-QAM.

\begin{figure*}[t]
	\centering
	\begin{subfigure}{0.32\linewidth}
		\centering
		\includegraphics[width=1\linewidth]{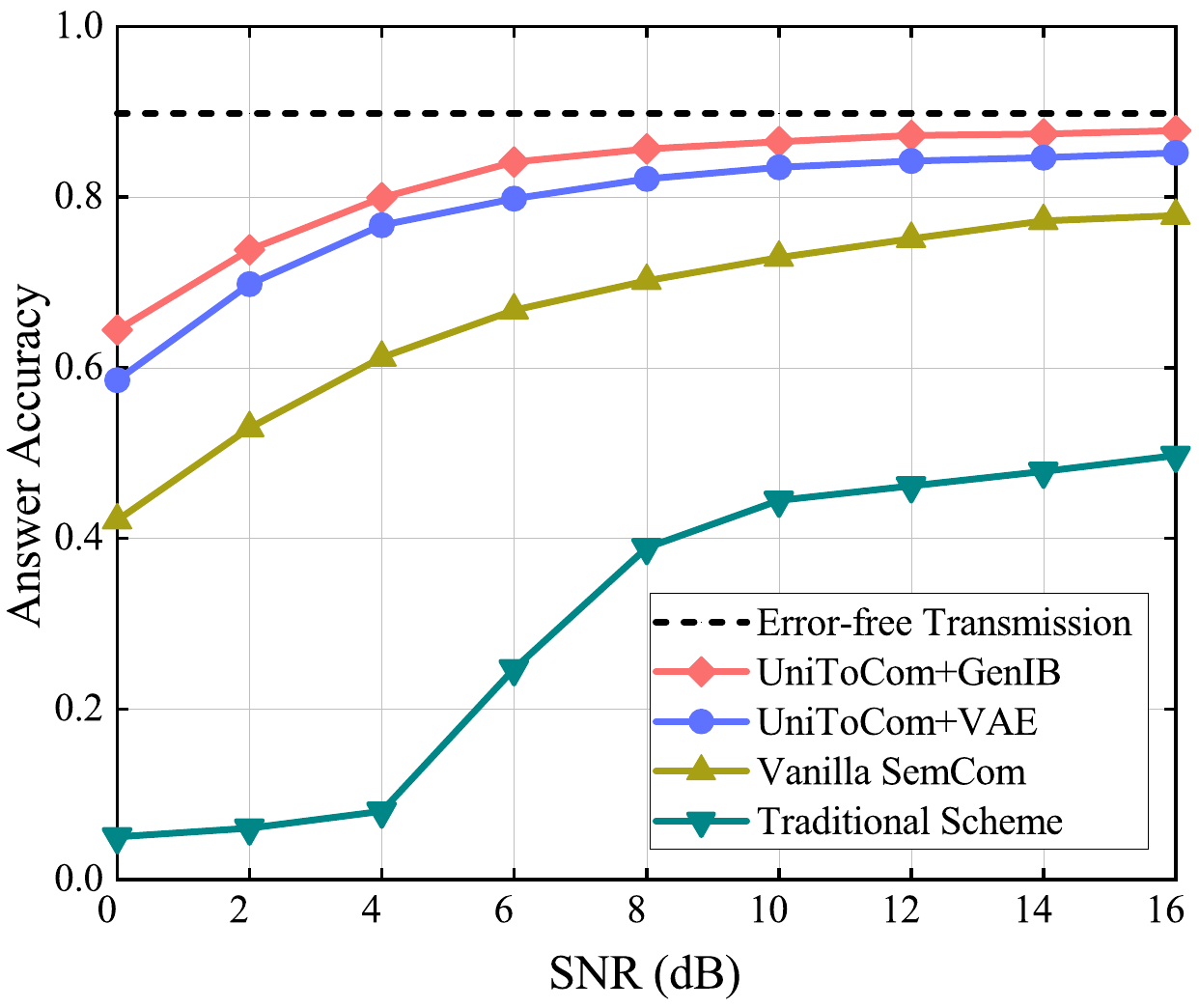}
		\caption{VQA understanding task}
		\label{Rayleigh-VQA}
	\end{subfigure}
	\begin{subfigure}{0.32\linewidth}
		\centering
		\includegraphics[width=1\linewidth]{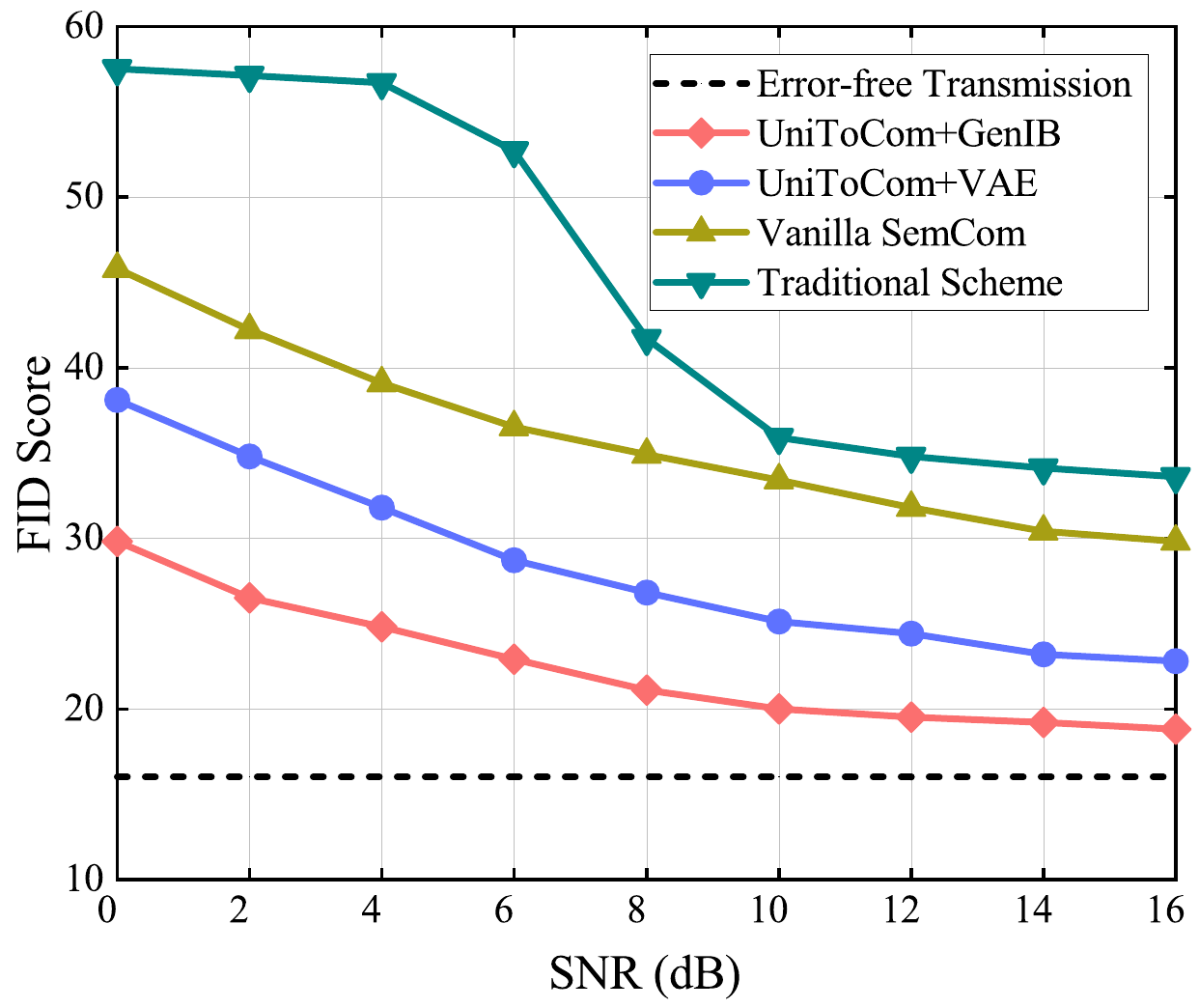}
		\caption{Text-to-image generation task}
		\label{Rayleigh-T2I}
	\end{subfigure}
	\begin{subfigure}{0.32\linewidth}
		\centering
		\includegraphics[width=1\linewidth]{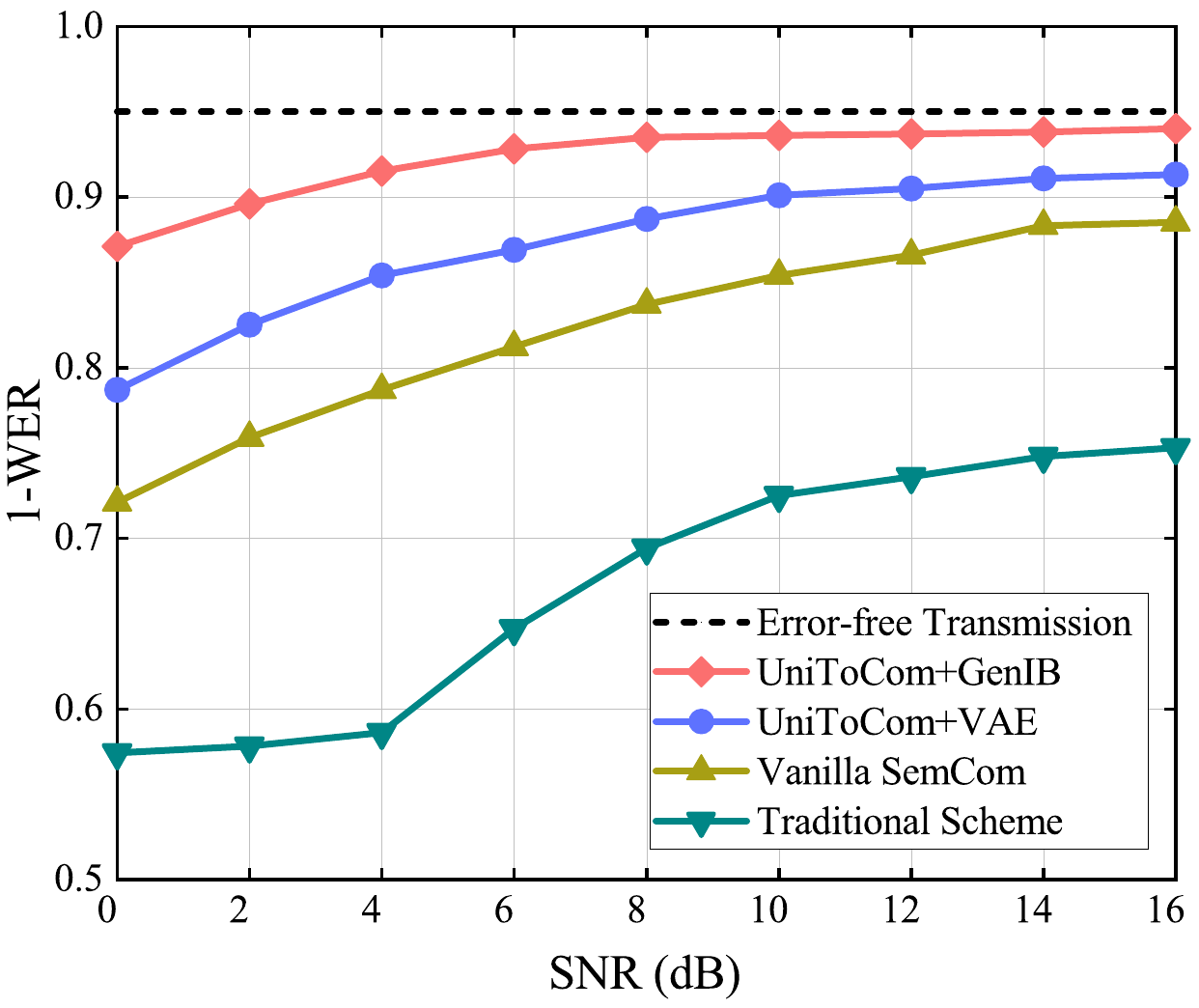}
		\caption{ASR generation task}
		\label{Rayleigh-ASR}
	\end{subfigure}
	\caption{Model performance versus SNR for various tasks under Rayleigh channels.}
	\label{Rayleigh-tasks}
\end{figure*}

Fig. \ref{Rayleigh-tasks} presents the performance comparison across different tasks under varying signal-to-noise ratio (SNR) regimes over Rayleigh fading channels.
It is evident that both UniToCom and vanilla SemCom significantly outperform traditional methods, which 
fail to effectively combat channel impairments, particularly in low-SNR scenarios.
Besides, 
UniToCom approaches the performance upper bound in high-SNR regimes, indicating its robustness and near-optimality under favorable channel conditions.
In contrast to the vanilla SemCom, our proposed UniToCom consistently achieves better accuracy and semantic fidelity across all evaluated tasks using a single unified model empowered by LLM.
This highlights its superior capability in jointly handling multimodal understanding and generation tasks within a unified communication paradigm.

As shown in Fig. 2(a), our proposed $\sigma$-GenIB-based UniToCom is able to achieve a similar accuracy compared to the VAE-based tokenization.
However, for generation-intensive tasks, UniToCom+GenIB obtains lower restoration distortion and higher generation quality as depicted in Fig. 2(b) and Fig. 2(c).
This performance gap underscores the critical role of GenIB in predictive accuracy and generative expressiveness.
Furthermore, the integration of an MLLM contribute significantly to UniToCom’s ability to seamlessly process heterogeneous modalities under diverse communication scenarios.

\begin{figure}[t]
	\centering
	\includegraphics[width=2.6in]{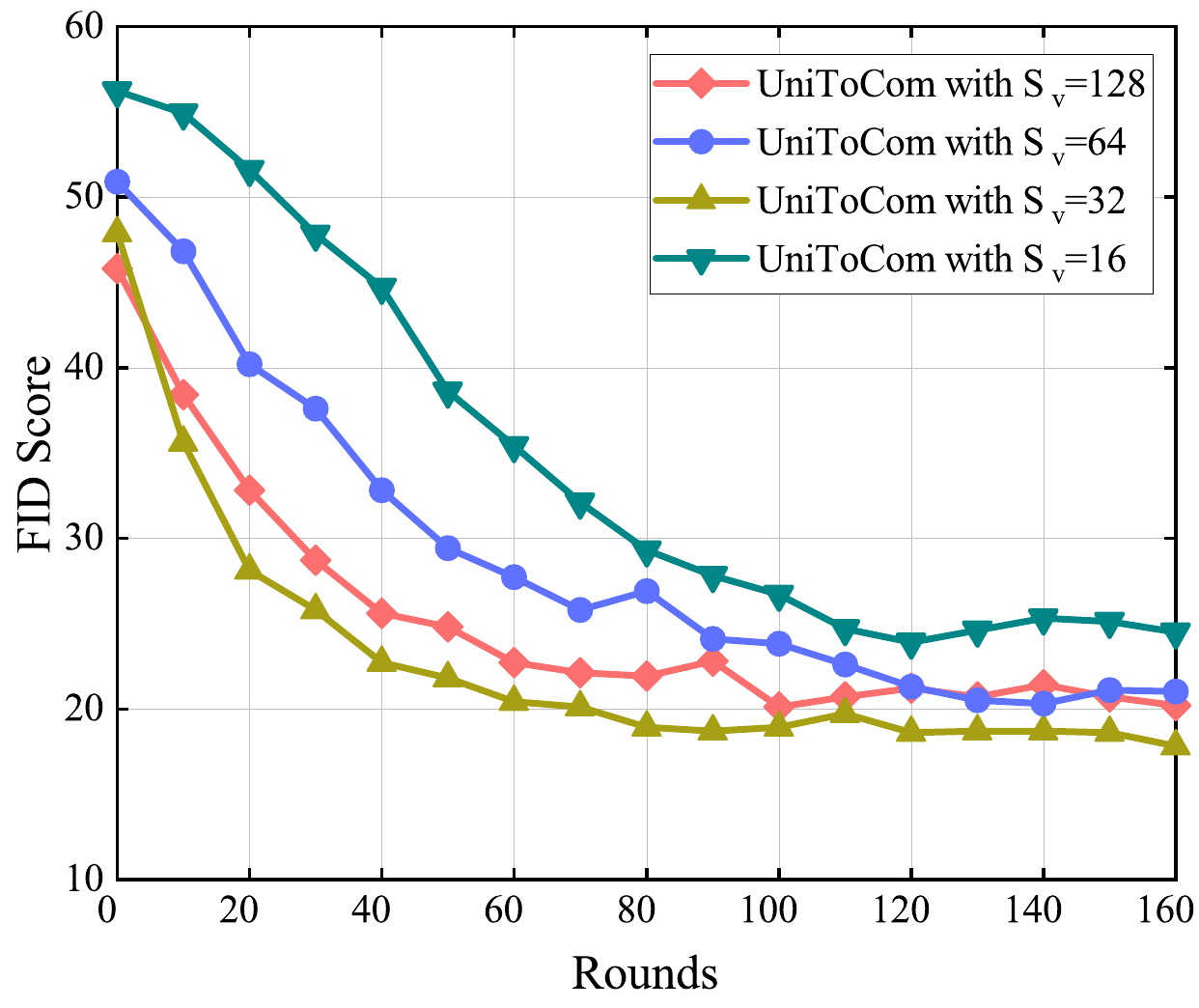}
	\caption{Performance comparisons of image token length.}
	\label{token-length}
\end{figure}

Fig. \ref{token-length} depicts the influence of different image token lengths.
The results indicate that, 
for a given task, 
appropriately compressing the token length can yield improved performance. Notably, the optimal performance is achieved when the image token length is set to 32, rather than the maximum length of 128. This suggests that excessively long token sequences may contain redundant information, which can introduce noise and impair model accuracy. Conversely, while overly aggressive compression ($S_v=16$) leads to a significant loss of essential information and degraded performance, moderate compression strikes a balance between information preservation and representational compactness. Furthermore, reducing token length substantially lowers the input dimensionality to the MLLM backbone, which in turn accelerates convergence and enhances both communication and computational efficiency.

\section{Conclusion}
In this letter, we have presented UniToCom, which treated tokens as the processible  transmission units.
Based on the proposed GenIB principle, UniToCom enabled the learning of compressed yet informative token representations. 
Then, we have developed $\sigma$-GenIB to address the issue of variance collapse  and preserve the diversity and stability of latent tokens. Moreover, the integration of an MLLM has facilitated unified processing of discrete and continuous tokens.
Simulation results have verified the superiority of UniToCom 
for multimodal understanding and generation.
Besides, moderate token compression can effectively reduce redundancy and accelerates convergence, which improved both communication and computational efficiency of wireless networks in the era of large models.

\end{document}